\documentclass{iopart}
\def\RR{{\rm I\kern-.1567em R}}                              
 \def\CC{{\rm C\kern-4.7pt                                    
 \vrule height 7.7pt width 0.4pt depth -0.5pt \phantom {.}}}
 \def\ZZ{{\sf Z\kern-4.5pt Z}}

\begin{document}

\title[Integrability in $U(1)$ Gauge Theories]
{Integrability in Theories with local $U(1)$ Gauge
Symmetry}

\author{C. Adam$^1$, J.
S\'{a}nchez-Guill\'{e}n$^1$ and A. Wereszczy\'{n}ski$^2$}

\address{$^1$ Departamento de Fisica de Particulas, Universidad
       de Santiago and Instituto Galego de Fisica de Altas Enerxias
       (IGFAE) E-15782 Santiago de Compostela, Spain}

\address{$^2$ Institute of Physics,  Jagiellonian University,
       Reymonta 4, 30-059 Krak\'{o}w, Poland}

\eads{ \mailto{adam@fpaxp1.usc.es} \mailto{joaquin@fpaxp1.usc.es}
\mailto{wereszczynski@th.if.uj.edu.pl}}

\begin{abstract}
Using a recently developed method, based on a generalization of
the zero curvature representation of Zakharov and Shabat, we study
the integrability structure in the Abelian Higgs model. It is
shown that the model contains integrable sectors, where
integrability is understood as the existence of infinitely many
conserved currents. In particular, a gauge invariant description
of the weak and strong integrable sectors is provided. The
pertinent integrability conditions are given by a $U(1)$
generalization of the standard strong and weak constraints for
models with two dimensional target space. The Bogomolny sector is
discussed, as well, and we find that each Bogomolny configuration
supports infinitely many conserved currents. Finally, other models
with $U(1)$ gauge symmetry are investigated.
\end{abstract}
\section{Introduction}
One of the most important features of realistic field theories, both in
elementary particle theory and in applications to condensed matter systems,
is their nonlinearity. Unfortunately, nonlinear field theories are
notoriously difficult to analyse. In low dimensional base spaces, the
concept of integrability, i.e., the existence of infinitely many
conserved quantities, has proven very useful for the investigation
of nonlinear field theories. In higher dimensions, much less
is known about nonlinear theories, and a general concept of integrability
has not yet been developed there. However, a generalization of the
zero-curvature condition of Zakharov and Shabat has been proposed in
\cite{integrability1}, and has been used to construct
non-linear field theories which have
either infinitely many conservation laws in the full theory, or which contain
integrable subsectors defined by some additional constraint equations on the
fields.
Applications of this construction have shown  that usually
these conserved currents
in the models or their subsectors are Noether currents
and generalizations thereof, i.e.,
they are related to
the geometry and symmetries of the target space (see, e.g., \cite{BF1}). So a
direct, geometric approach has been successfully undertaken to find
these conservation laws, e.g.,  for models with two- and
three-dimensional target space (like, e.g.,  the
  Faddeev--Niemi and Skyrme models), \cite{chiral}, \cite{skyrme1},
\cite{ASG1},
\cite{weak}, \cite{ASGW-Vol}. Generalization to higher dimensional
target space ($CP^n$ model) has been established, as well \cite{cpn
1}, \cite{skyrme-cpn2}. So far this method has only been used in
the investigation of field theories without gauge symmetries. The
main aim of the present work is to adopt this approach to some
nonlinear field theories with an Abelian gauge symmetry. In
particular, we will perform calculations for the Abelian Higgs
model. It is worth underlining that our analysis is valid for an
arbitrary, $d+1$ dimensional space-time.
\\
Obviously, the main problem in the application of the zero
curvature framework to theories with a local gauge symmetry is
the gauge invariance of the integrable
subsectors. Fortunately we find that one can define such
sectors in a completely gauge independent manner.
\\
The paper is organized as follows. In Section 2 we briefly recall the
integrability of a class of non-gauge theories for later convenience.
Then we construct the integrability conditions and the corresponding conserved
currents for the Abelian Higgs model. Further, we discuss the relation of
the integrable sectors to the Bogomolny sector of the model.
In Section 3 we discuss some
generalizations. Section 4 contains our conclusions.
\section{Integrability in the Abelian Higgs model}
The Abelian Higgs model is given by the following Lagrangian density
\begin{equation}
{\cal L} =-F^2_{\mu \nu}  + \frac{1}{2} (D_{\mu}u) (D^{\mu}u)^* -V,
\end{equation}
where the covariant derivative is $D_{\mu} u = u_{\mu} -ieA_{\mu}u$ and
$u_\mu \equiv \partial_\mu u$.
Moreover, $F_{\mu \nu}$ is the standard antisymmetric field tensor
for the Abelian, real gauge field $A_{\mu}$. The Higgs potential
reads
\begin{equation}
V=\frac{\lambda}{4} \left( \frac{m^2}{\lambda} - uu^* \right)^2.
\end{equation}
Here, $e$ and $\lambda$ are gauge and scalar coupling constants
whereas $m$ fixes the vacuum value of the Higgs field.
\\
The canonical momenta are
\begin{equation}
\pi_{\mu} \equiv \frac{\partial L}{\partial u_{\mu}} = \frac{1}{2}
(D_{\mu} u )^*, \;\;\; \pi_{\mu}^* \equiv \frac{\partial L}{\partial
u_{\mu}^*}= \frac{1}{2} D_{\mu} u
\end{equation}
and
\begin{equation}
\pi_{\mu \nu} \equiv \frac{\partial L}{\partial
(\partial^{\mu}A^{\nu})}=-4F_{\mu \nu}.
\end{equation}
The pertinent equations of motion read
\begin{equation}
\partial_{\mu} \pi^{\mu}=L_u=-ie A_{\mu}\pi^{\mu}-V'u^* , \;\;\;
\partial_{\mu}
\pi^{*\mu}=L_{u^*}=ieA_{\mu}\pi^{*\mu}-V'u,
\end{equation}
\begin{equation}
\partial_{\mu}F^{\mu\nu}= \frac{ie}{8} \left[ (D^{\nu}
u)^*u-(D^{\nu}u)u^* \right], \label{eq mot gauge}
\end{equation}
where the prime denotes differentiation with respect to $uu^*$.
\subsection{The non-gauge case}
Before defining the integrable sectors and the conserved currents in the
Abelian Higgs model, we find it convenient to briefly recall the analogous
situation for some non-gauge models. Integrability for a large class of
theories with two-dimensional target space given by a complex field $u$
has been discussed in \cite{ASG1}, \cite{weak}. Here we just want to quote
results for a small subclass of models, which are useful for our purposes.
For the class of CP$^1$ type models
\begin{equation} \label{cp1-model}
{\cal L}_{\rm CP^1} = f(uu^*) u^\mu u^*_\mu
\end{equation}
the currents
\begin{equation} \label{strong-cu}
j_\mu = \frac{i}{f}(G_u \pi^*_\mu - G_{u^*} \pi_\mu )
\end{equation}
are conserved provided that the additional constraint or integrability
condition (the complex eikonal equation)
\begin{equation}
u^\mu u_\mu =0
\end{equation}
holds.  Here $\pi_\mu$ is the canonical four-momentum of the CP$^1$ type model
(\ref{cp1-model}), and $G\equiv G(u,u^*)$ is an arbitrary real function
of the field $u$ and its complex conjugate; further, $G_u \equiv \partial_u
G$.
We call this condition the strong
integrability condition and the corresponding sector of fields the strong
integrable sector.
\\
For the class of  CP$^1$ type models with an additional potential
\begin{equation} \label{cp1-model-pot}
{\cal L}_{\rm CP^1,V} = f(uu^*) u^\mu u^*_\mu - V(uu^*)
\end{equation}
the currents
\begin{equation} \label{weak-cu}
j_{\mu} =i (G_u \pi^*_\mu - G_{u^*} \pi_\mu ) = iG' (u^* \pi^*_\mu
- u \pi_\mu )
\end{equation}
are conserved provided that the additional constraint or integrability
condition (the weak integrability condition)
\begin{equation}
u^{*2} u^\mu u_\mu - u^2 u^{*\mu}u^*_\mu =0
\end{equation}
holds. Here it is assumed that $G\equiv G(uu^*)$ is a function of the modulus
squared $uu^*$ only. Observe that the factor $f^{-1}$, which is present in
the definition of the conserved currents of the strong sector, is unnecessary
for the currents of the weak sector, because it can be compensated by a
redefinition of the arbitrary $G(uu^*)$.
\\
How can we expect that these results carry over to the case of a gauge theory,
specifically to the Abelian Higgs model?
Gauge invariant generalizations of the weak and strong integrability
conditions are naturally given by
\begin{equation} \label{int cond weak}
u^2(D_{\mu}u)^{*2}-u^{*2}(D_{\mu}u)^2=0
\end{equation}
(the weak gauge invariant condition), and by
\begin{equation} \label{int cond strong}
(D_{\mu}u)^2=0
\end{equation}
(the strong gauge invariant condition), as we shall obtain in the sequel.
We will find, however, that there exists a subtle
difference between the weak and strong sectors concerning the currents. The
weak currents (\ref{weak-cu}), when defined for the Abelian Higgs model,
are gauge invariant (because of $G=G(uu^*)$) and
have gauge invariant conservation equations. Their conservation leads
therefore directly to the weak condition (\ref{int cond weak}), as we shall
see. On the other
hand, the currents (\ref{strong-cu}) for the Abelian Higgs model
are not gauge invariant and do not
give rise to gauge invariant conservation equations (because of
$G=G(u,u^*)$). Therefore, the strong condition (\ref{int cond strong}) has
to be interpreted differently. We will find that it
means the {\em existence} of a gauge such
that the strong currents (\ref{strong-cu}) are conserved in this gauge. It is
this latter condition which is gauge invariant.
\\
We have indicated, right now, the existence of conserved currents in subsectors
of the Abelian Higgs model, which are
in close analogy with the case of non-gauge theories. One might ask whether
there exist further conserved currents in some integrable sectors,
which have no
analogy in non-gauge theories. Interestingly, the answer to this question is
yes, as we shall see shortly.
\subsection{The weak integrable gauge sector}
Taking into account the above discussion, we assume the following family
of currents for the Abelian Higgs model
\begin{equation}
j_{\mu}=i(G_u \pi_{\mu}^* - G_{u^*} \pi_{\mu}), \label{currents}
\end{equation}
where $G$ is an arbitrary function of the modulus $|u|$ squared, that is
$G=G(uu^*)$. These currents are conserved if
\begin{eqnarray}
0=-i \partial_{\mu}j^{\mu}&=& G''(uu^*_{\mu}+u^*u_{\mu})(u^*\pi^{*\mu}
-u\pi^{\mu}) \nonumber \\
&&
+G'(u^*_{\mu}\pi^{*\mu}-u_{\mu}
\pi^{\mu} +u^*\partial^{\mu} \pi_{\mu}^*-u
\partial^{\mu}\pi_{\mu}).
\end{eqnarray}
After some calculations one finds that the condition providing
conservation of the currents takes the form
\begin{equation}
0=u^{*2} u_{\mu}^2 - u^2 u^{*2}_{\mu} -2ieuu^*[ u u^*_{\mu} +u^*
u_{\mu}] A^{\mu}. \label{int cond weak1}
\end{equation}
It is easy to notice that this equation is the natural generalization of the
well-known weak integrable condition for models with two
dimensional target space \cite{weak}. Therefore we  call this
sector of the Abelian Higgs model, defined by the additional constraint
(\ref{int cond weak1}), the weak one. The weak integrability
condition reveals its geometrical meaning if we re-express it in terms
of the polar decomposition of the Higgs field
\begin{equation}
u=e^{\Sigma+i\Lambda},
\end{equation}
where $\Sigma, \Lambda$ are real fields. Then we get
\begin{equation}
\Sigma_{\mu} (\Lambda^{\mu} - e A^{\mu})=0. \label{int cond weak2}
\end{equation}
Thus the weak integrable sector is defined by field configurations
such that the gradient of the modulus is perpendicular to a covariant
derivative of the phase.
\\
As the full theory possesses the $U(1)$ gauge symmetry, we should
also have a gauge invariant description for the integrable sector.
One can check that our constraint as well as the currents are
invariant under the gauge transformation
\begin{equation}
\Sigma \rightarrow \Sigma, \;\;\; \Lambda \rightarrow \Lambda +ef,
\;\;\;A_{\mu} \rightarrow A_{\mu} +\partial_{\mu} f.
\end{equation}
In fact, we can rewrite expression (\ref{int cond weak1}) in an
elegant, manifestly gauge independent way as
$u^2(D_{\mu}u)^{*2}-u^{*2}(D_{\mu}u)^2=0$, as announced by
Eq. (\ref{int cond weak}).
As said, this is the $U(1)$ gauge invariant generalization of the standard
weak constraint.
\\
Of course, one should ask whether such a submodel has any solution
at all. The answer is yes. For example, let us consider the
standard static, axially symmetric vortex Ansatz in 2+1 dimensions
\cite{nielsen},
\begin{equation}
\fl u(\rho, \phi)=e^{in \phi} v (\rho), \;\;\; A_0=0, \;\;
A_1=-\frac{y}{\rho} A(\rho), \;\; A_2=\frac{x}{\rho}A(\rho),
\end{equation}
where $\rho=\sqrt{x^2+y^2}$ and the functions $v(\rho), A(\rho)$
still have to be determined from the pertinent ordinary second
order differential equations (a trivial immersion into arbitrary
$d > 2$ space dimensions shows that the pertinent weak sector in
higher dimensions is non-empty as well). Such solutions of the
Abelian Higgs model with a non-critical coupling are known
numerically or in an approximated form \cite{arodz},
\cite{karkowski}. As we see, for this Ansatz
\begin{equation}
\Lambda = n\phi, \;\; \Sigma = \ln v (\rho).
\end{equation}
Obviously these fields obey the weak integrability condition. This
fact resembles the results obtained for $CP^n$ model, where
non-BPS solutions belong to one of the weak sectors
\cite{skyrme-cpn2}.
\\
It should be underlined that the weak integrable condition gives
the conserved currents for all values of the scalar coupling
constant $\lambda$.
\\
Finally let us point out that the currents are conserved {\it half
off-shell}. Indeed, the conservation law is fulfilled if the
equations of motion for the Higgs field are imposed. On the other
hand, it is not necessary to assume that the gauge field satisfies
(\ref{eq mot gauge}).
\subsection{New currents in the weak integrable sector}
Rather surprisingly, we are able to construct another family of
conserved currents in the weak sector. Namely,
\begin{equation}
j_{\mu}=F_{\mu \nu} \left( H_u \pi^{*\nu} + H_{u^*}\pi^{\nu}
\right), \label{currents new}
\end{equation}
where the arbitrary function $H$ depends on the modulus $|u|$ squared, i.e.,
$H=H(uu^*)$.
\\
Then
\begin{eqnarray}
\partial_{\mu} j^{\mu}= \frac{H'}{2} \partial_{\mu} F^{\mu \nu}
( u^* u^{\nu} + u
u^{*\nu})+ & \nonumber \\ \fl \frac{1}{2}F_{\mu \nu} \left[ H' (u^{* \mu}
u^{\nu} + u^* u^{\mu \nu} + u^{\mu} u^{*\nu}+ u u^{*\mu \nu}
\right) + H'' (u^{\mu} u^* + u u^{*\mu} )( u^* u^{\nu} + u
u^{*\nu})].
\end{eqnarray}
We easily get
\begin{equation}
\partial_{\mu} j^{\mu}= \frac{ieH'}{16} [ (D^{\nu}
u)^*u-(D^{\nu}u)u^* ] \; [ u^* u^{\nu} + u u^{*\nu}].
\end{equation}
or after the polar decomposition
\begin{equation}
\partial_{\mu} j^{\mu}= \frac{H' (uu^*)^2}{4} ( \Lambda_{\nu}
-eA_{\nu} ) \Sigma^{\nu}.
\end{equation}
Due to the fact that the fields belong to the weak integrable
sector this expression vanishes.
\\
Similarly as before this new class of conserved currents is
gauge independent and exists for any scalar coupling constant.
Moreover, they are also conserved {\it half off-shell}. However,
inversely to the previously discussed case, one must impose
the equation of motion for the gauge field while the complex field
does not have to obey the field equations.
\subsection{The strong integrable sector}
In order to find
 more conserved currents, we now set
$\lambda=0$, i.e., we neglect the potential term in the Lagrangian,
analogously to the non-gauge case discussed in Section 2.1.
\\
The currents are given by the same expression as in the weak case,
\begin{equation}
j_{\mu}=i(G_u \pi_{\mu}^* - G_{u^*} \pi_{\mu}), \label{currents
strong}
\end{equation}
but now the function $G$ depends on $u$ and $u^*$ in a completely
arbitrary manner, $G=G(u,u^*)$.
\\
Then
\begin{eqnarray}
-i \partial_{\mu}j^{\mu} &=& G_{uu}u_{\mu}\pi^{*\mu} + G_{uu^*}
u^*_{\mu} \pi^{*\mu} -G_{u^*u}u_{\mu}\pi^{\mu}
-G_{u^*u^*}u^*_{\mu} \pi^{\mu} \nonumber \\
&&
+ G_u\partial_{\mu}\pi^{*\mu} -G_{u^*}\partial_{\mu} \pi^{\mu}
\end{eqnarray}
vanishes if
\begin{equation}
u^*_{\mu} \pi^{*\mu} -u_{\mu}\pi^{\mu}=0, \;\; \;
 \label{strong gen cd1}
\end{equation}
\begin{equation}
u_{\mu} \pi^{*\mu}=0, \;\; \; u_{\mu}^* \pi^{\mu}=0, \label{strong
gen cd2}
\end{equation}
\begin{equation}
A_{\mu} \pi^{*\mu}=0, \;\; \; A_{\mu} \pi^{\mu}=0 \label{strong
gen cd3}
\end{equation}
or, if expressed using the polar decomposition,
\begin{equation}
\Sigma_{\mu}A^{\mu}=0, \label{str cond 1}
\end{equation}
\begin{equation}
\Sigma_{\mu}\Lambda^{\mu}=0, \label{str cond 2}
\end{equation}
\begin{equation}
\Sigma_{\mu}^2 -\Lambda_{\mu}^2+e\Lambda_{\mu} A^{\mu}=0,
\label{str cond 3}
\end{equation}
\begin{equation}
A_{\mu}(\Lambda^{\mu} -eA^{\mu})=0. \label{str cond 4}
\end{equation}
We see immediately that these so-called strong
integrability conditions are gauge dependent.
\\
To make sense of the strong integrable sector, we have to
understand integrability in gauge models as {\it the existence of
a gauge in which there are infinitely many conserved currents},
that is {\it the existence of a gauge in which conditions
(\ref{str cond 1})-(\ref{str cond 4}) are fulfilled}. Then we
may ask another question, namely {\it what should be the conditions for
these fields which would provide the existence of such a gauge}.
\\
So, let us take arbitrary fields $\Sigma',\Lambda',A_{\mu}'$. In
the most general case the expressions for the strong constraints
(\ref{str cond 1})-(\ref{str cond 4}) are nonzero and equal to
\begin{equation}
\Sigma'_{\mu} \Lambda'^{\mu}=g, \;\;\;\; \Sigma'_{\mu} A'^{\mu}=h,
\end{equation}
\begin{equation}
\Sigma_{\mu}'^2 -\Lambda_{\mu}'^2+e\Lambda'_{\mu} A'^{\mu}=k, \;\;
A'_{\mu}(\Lambda'^{\mu} -eA'^{\mu})=j,
\end{equation}
where $g,h,j,k$ are arbitrary functions. Using the gauge
transformations we express these fields by new ones
\begin{equation}
\Sigma'=\Sigma, \;\; \Lambda'=\Lambda+ef, \;\;
A'_{\mu}=A_{\mu}+f_{\mu}.
\end{equation}
Thus
\begin{equation}
\Sigma_{\mu} \Lambda^{\mu}+ ef^{\mu} \Sigma_{\mu}=g, \;\;\;\;
\Sigma_{\mu} A^{\mu}+ f^{\mu}\Sigma_{\mu}=h,
\end{equation}
\begin{equation}
\Sigma_{\mu}^2 -\Lambda_{\mu}^2+e\Lambda_{\mu}
A^{\mu}+ef^{\mu}(eA_{\mu}-\Lambda_{\mu})=k,
\end{equation}
\begin{equation}
A_{\mu}(\Lambda^{\mu} -eA^{\mu})+f^{\mu}(\Lambda_{\mu}-
eA_{\mu})=j.
\end{equation}
Additionally, the new fields are assumed to obey the strong
conditions. Then we find
\begin{equation}
ef^{\mu} \Sigma_{\mu}=g, \;\;\; f^{\mu}\Sigma_{\mu}=h,
\end{equation}
\begin{equation}
ef^{\mu}(eA_{\mu}-\Lambda_{\mu})=k, \;\;\; f^{\mu}(\Lambda_{\mu}-
eA_{\mu})=j.
\end{equation}
These formulas are not self-contradictory if
\begin{equation}
eh=g \;\; \wedge \;\; -ej=k.
\end{equation}
It means that
\begin{equation}
\Sigma_{\mu}(\Lambda^{\mu}-eA^{\mu})=0, \label{strong c1}
\end{equation}
\begin{equation}
\Sigma_{\mu}^2=(\Lambda_{\mu}-eA_{\mu})^2. \label{strong c2}
\end{equation}
These conditions are gauge independent and can be rewritten in an
elegant form like $
(D_{\mu}u)^2=0, $ as announced in Eq. (\ref{int cond strong}),
which is nothing else but the $U(1)$ generalization of the strong
integrability condition known for models with two dimensional
target space.
\\
Of course, the strong integrability condition (\ref{int cond strong})
can be defined for all values of
the coupling constant $\lambda$. Nonetheless, except in the case when
$\lambda=0$, it does not give rise to more conserved currents than
in the weak sector.
\\
The fields which obey (\ref{int cond strong}), obey (\ref{int cond
weak}) as well. Therefore, the strong sector is a subset of the
weak sector.
\subsection{The Bogomolny sector}
It is a well-known fact that for a special value of the
coupling $\lambda=1$ (we set $e=m=1$), there is a Bogomolny
sector in the Abelian Higgs model \cite{bogom} in 2+1 dimensions.
This sector is
defined by the static, first order differential equations
\begin{equation}
F_{12}+\frac{1}{2} (|u|^2-1)=0, \label{bogom 1}
\end{equation}
\begin{equation}
(D_1u + iD_2u)=0, \label{bogom 2}
\end{equation}
where we assumed $A_0=0$. Static, multi-vortex solutions of these
equations as well as their dynamics have been investigated
by many authors \cite{bogom}-\cite{gonzalez2}.
\\
Let us now discuss the Bogomolny sector and its role from the
integrability point of view.
\\
First of all, the second Bogomolny equation (\ref{bogom 2}) is a
square root of the strong integrability condition. In fact, for
static configurations we have
\begin{equation}
\fl
-(D_{\mu}u)^2=(D_iu)^2=(D_1u)^2+(D_2u)^2=(D_1u+iD_2u)(D_1-iD_2u).
\end{equation}
Thus, the Bogomolny sector is a subset of the strong as well as
weak integrable sector. Moreover, after the polar decomposition
formula (\ref{bogom 2}) reads
\begin{equation}
\Sigma_1-\Lambda_2+eA_2=0, \;\;\; \Sigma_2+\Lambda_1-eA_1=0,
\end{equation}
i.e.,
\begin{equation}
\Lambda_i-eA_i=-\epsilon_{ij} \Sigma_{j}, \;\;\; i,j=1,2.
\end{equation}
It is worth stressing that the Bogomolny sector supports the
existence of an infinite set of non-trivial conservation laws as
conjectured by Maison \cite{maison}. As the coupling constant is
not zero, only the weak currents given by formula (\ref{currents}) are
conserved. Then, the non-zero components
\begin{equation}
j_k=iG'(uu^*) uu^* (\Lambda_k -eA_k)
\end{equation}
take the form
\begin{equation}
j_k=-iG'(e^{2 \Sigma}) e^{2 \Sigma} \epsilon_{kl} \Sigma_l.
\label{bps currents}
\end{equation}
In accordance with our former discussion, the currents are
identically conserved for an arbitrary function $\Sigma$.
\\
An analogous analysis can be performed for the new weak currents of Section
2.3. We get
\begin{equation}
j_k= \frac{H'}{2} F_{kl} \Sigma_l.
\end{equation}
In this case we use the first Bogomolny equation and obtain
\begin{equation}
j_k= -\frac{H'}{4} (e^{2\Sigma}-1) \epsilon_{kl} \Sigma_l.
\end{equation}
As we see, because of the arbitrariness of the functions $G$ and
$H$, both types of currents lead to identical expressions for the
Bogomolny configurations. In other words, in the Bogomolny sector
the two families of currents in the weak sector are
undistinguishable.
\section{Generalizations}
It is straightforward to generalize our results to other models
with a $U(1)$ gauge symmetry. Some specific examples are discussed
below.
\subsection{Generalized $U(1)$ CP$^1$ model: type I}
A possible generalization of $U(1)$ CP$^1$ model is provided by
assuming a more general form of the kinetic part for the scalar
field. Namely,
\begin{equation}
L=-F^2_{\mu \nu} + f(uu^*)(D_{\mu} u )(D^{\mu}u)^*-V(uu^*).
\end{equation}
One can prove that the weak as well as the strong sectors are defined
via the same formulas as in the previous section. The only
difference is in the form of the strong currents which should be
\begin{equation}
j_{\mu}=\frac{i}{f(uu^*)} (G_u\pi_{\mu}^*-G_{u^*}\pi_{\mu})
\end{equation}
(see the remark in Section 2.1). As the most prominent example of
this class of Lagrangians let us mention the gauged $O(3)$ sigma
model \cite{schroers}, for which $f=V=4/(1+|u|^2)^2$.
\subsection{Generalized $U(1)$ CP$^1$ model: type II}
In this case the Lagrangian is given by the formula \cite{lee}
\begin{equation}
L=-g(uu^*)F^2_{\mu \nu} + \frac{1}{2}(D_{\mu} u
)(D^{\mu}u)^*-V(uu^*). \label{gen ah 2}
\end{equation}
The weak sector and currents remain unchanged. However, there are
no new currents in the strong sector, even if we neglect the
potential $V$. Indeed, the first term in (\ref{gen ah 2})
effectively plays the role of a potential for the scalar field. The
pertinent Bogomolny solutions (if they exist) form a subset of the
weak sector.
\subsection{Abelian Chern-Simons-Higgs model}
Analogously, the same integrable
structure appears in the Abelian Chern-Simons-Higgs model
\cite{csh1}-\cite{ferreira}
\begin{equation}
L=\frac{\kappa}{4}\epsilon^{\alpha \beta \gamma} A_{\alpha}
F_{\beta \gamma} +\frac{1}{2} (D_{\mu} u)^* (D^{\mu} u) -
\frac{\lambda}{8} |u|^2(|u|^2-v^2)^2.
\end{equation}
Once again, the conditions for the Bogomolny sector imply the strong
integrability conditions, and the strong conditions imply the weak ones.
The Bogomolny sector exists only
for the critical value of the coupling, where
$\lambda=\frac{e^4}{\kappa^2}$.
\subsection{Model with a non-minimal coupling}
Finally, let us comment on a model with a non-minimal coupling
between the scalar and gauge field, where the scalar field is not
gauged,
\begin{equation}
L=-\frac{1}{4} g(uu^*) F^{\mu \nu}  F_{\mu \nu} +\frac{1}{2}
u_{\mu}^* u^{\mu} - V(uu^*).
\end{equation}
The integrable structure remains unchanged. However, there are two
families of new currents which are conserved if the equation
of motion for the gauge field is obeyed.
\begin{equation}
j^{(1)}_{\mu}= H^{(1)}(uu^*) g F_{\mu \nu} \left[ u^*u^{\nu} +
uu^{*\nu} \right],
\end{equation}
\begin{equation}
j^{(2)}_{\mu} = H^{(2)} \left( \frac{u}{u^*} \right) g F_{\mu \nu}
\left[ \frac{u^{\nu}}{u} - \frac{u^{*\nu}}{u^*} \right]
\end{equation}
i.e.
\begin{equation}
j^{(1)}_{\mu}= H^{(1)}(\Sigma ) g F_{\mu \nu} \Sigma^{\nu}, \;\;\;
j^{(2)}_{\mu}= H^{(2)} \left( \Lambda \right)  g F_{\mu \nu}
\Lambda^{\nu},
\end{equation}
where $H^{(1)}, H^{(2)}$ are arbitrary real functions of the
modulus and the phase respectively (the coupling function $g$ in
the first formula can be removed by a redefinition of the $H^{(1)}$
function). Notice that unlike for the models with local gauge
symmetry, no additional conditions have to be imposed to provide
the conservation of these currents.
Therefore, there should exist symmetries in this model such that
these conserved currents are their Noether currents (when additional
integrability conditions have to be imposed in order to have conserved
currents, this need not be true,
see \cite{ASG2}).
\section{Conclusions}
Let us briefly summarize the obtained results.
\\
The property of integrability, known from nonlinear
theories with global solitons, can be extended to models with
an Abelian gauge group. We have established the gauge invariant
description of the weak as well as strong integrable subsectors.
However, in spite of the fact that both integrability
conditions are gauge independent, only the weak currents do not
change under gauge transformations. This rather profound
difference might indicate that the weak integrable sector plays a
more important role than the strong one, at least for Lagrangians
with local gauge symmetry.
\\
On the other hand, the importance of the strong integrability
condition emerges from the observation that it is a Lorentz
invariant generalization of the Bogomolny equations.
\\
To conclude, the structure of the integrable sectors for all
models analyzed above is as follows
$$
 \mbox{Weak Sector} \;\; \supset \;\; \mbox{Strong
Sector} \;\; \supset \;\; \mbox{Bogomolny Sector},
$$
where the last inclusion makes sense only if the corresponding
Bogomolny sector exists. In the cases investigated in this paper
the Bogomolny sector only exists for values of the coupling
constants such that the strong sector and the Bogomolny sector do
not give rise to more conserved currents than the weak sector.
Still, the weak sector provides infinitely many conserved
currents, therefore we may specifically conclude that in the
Abelian Higgs model and related models each Bogomolny
configuration posesses infinitely many conserved currents.
\\
In cases when there is no Bogomolny sector (e.g., for noncritical values of
the couplings, or in higher dimensions),
the weak and strong integrable sectors still exist and provide
solutions with infinitely many conservation laws.
This may indicate
that in higher dimensions integrable submodels can play a role similar to
the Bogomolny sectors, providing a useful tool for the investigation of such
models and for the construction of solutions.

\ack C.A. and J.S.-G. thank MCyT (Spain) and FEDER
(FPA2005-01963), and support from Xunta de Galicia (grant
PGIDIT06PXIB296182PR and Conselleria de Educacion). A.W. is
partially supported from Adam Krzy\.{z}anowski Fund and
Jagiellonian University (grant WRBW 41/07). Further, C.A.
acknowledges support from the Austrian START award project
FWF-Y-137-TEC and from the FWF project P161 05 NO 5 of N.J.
Mauser.

\end{document}